# COMPUTATION OF INTERNAL FLUID FLOWS IN CHANNELS USING THE CFD SOFTWARE TOOL FLOWVISION

**ALEXEY N. KOCHEVSKY**
*Research Scientist, Department of Fluid Mechanics,*
*Sumy State University,*
*Rimsky-Korsakov str., 2, 40007, Sumy, Ukraine*
*alkochevsky@mail.ru*

**Abstract:** The article describes the CFD software tool FlowVision (OOO "Tesis", Moscow). The model equations used for this research are the set of Reynolds and continuity equations and equations of the standard $k - \varepsilon$ turbulence model. The aim of the paper was testing of FlowVision by comparing the computational results for a number of simple internal channel fluid flows with known experimental data. The test cases are non-swirling and swirling flows in pipes and diffusers, flows in stationary and rotating bends. Satisfactory correspondence of results was obtained both for flow patterns and respective quantitative values.

## 1. Introduction

Tendency of recent years is appearance and wide distribution of commercial CFD software tools allowing for performing of numerical computation of fluid flows of arbitrary complexity in the regions of arbitrary geometrical configuration. Dozens of such software tools are listed at the site www.cfd-online.com devoted to computational fluid dynamics. The number of publications in the leading international journals on fluid dynamics with results of successful application of commercial CFD tools for prediction of local and integral parameters of flows in hydraulic machines and other technical devices steadily grows (Ruprecht, 2002). Among the most popular CFD packages we should mention, in particular, CFX (www.software.aeat.com/cfx), STAR-CD (www.cd-adapco.com, www.adapco-online.com), Fluent (www.fluent.com), Numeca (www.numeca.be). Profound documentation is supplied with these and other CFD software tools, making it possible for a qualified enough person to use these tools successfully for fluid flow computations, with no or minor technical assistance from the developers. Of course, popularity of all these software tools is also advanced by proceeding growth of computational power of personal computers.

FlowVision (Aksenov et al., 1993) is one of the first commercial multi-purpose CFD software tools developed in Russia. This tool was granted for test application to the department of fluid mechanics of Sumy State University. In this paper, we present the results of our testing, by comparing the computation results obtained with this tool with known experimental data for a number of simple fluid flows in channels.

## 2. Description of CFD software tool FlowVision

The software tool FlowVision developed by OOO "Tesis" (Moscow, www.tesis.com.ru, www.flowvision.ru) is designed for numerical simulation of fluid flows and since the year 2000 is distributed as a commercial version. The results presented in the article are obtained using the release issued in February 2003. At the latter site, the full-functional demo-version of FlowVision is freely available (the sole restriction of the demo-version is that the number of nodes of the computational mesh can

not exceed 20 000).

The software tool FlowVision does not have own preprocessor but supports import of geometry from a number of modern CAD systems, e.g., SolidWorks. FlowVision has convenient interface permitting to examine the imported computational domain, specify boundary conditions, properties of the medium, computation parameters and perform other necessary operations.

The software tool FlowVision uses rectangular (Cartesian) computational mesh (Fig. 1). A user can arbitrarily thicken this mesh where necessary, e.g., near solid walls. Besides, a number of criteria are provided for automatic thickening of the mesh, e.g., in zones of rapid changing of computed flow parameters.

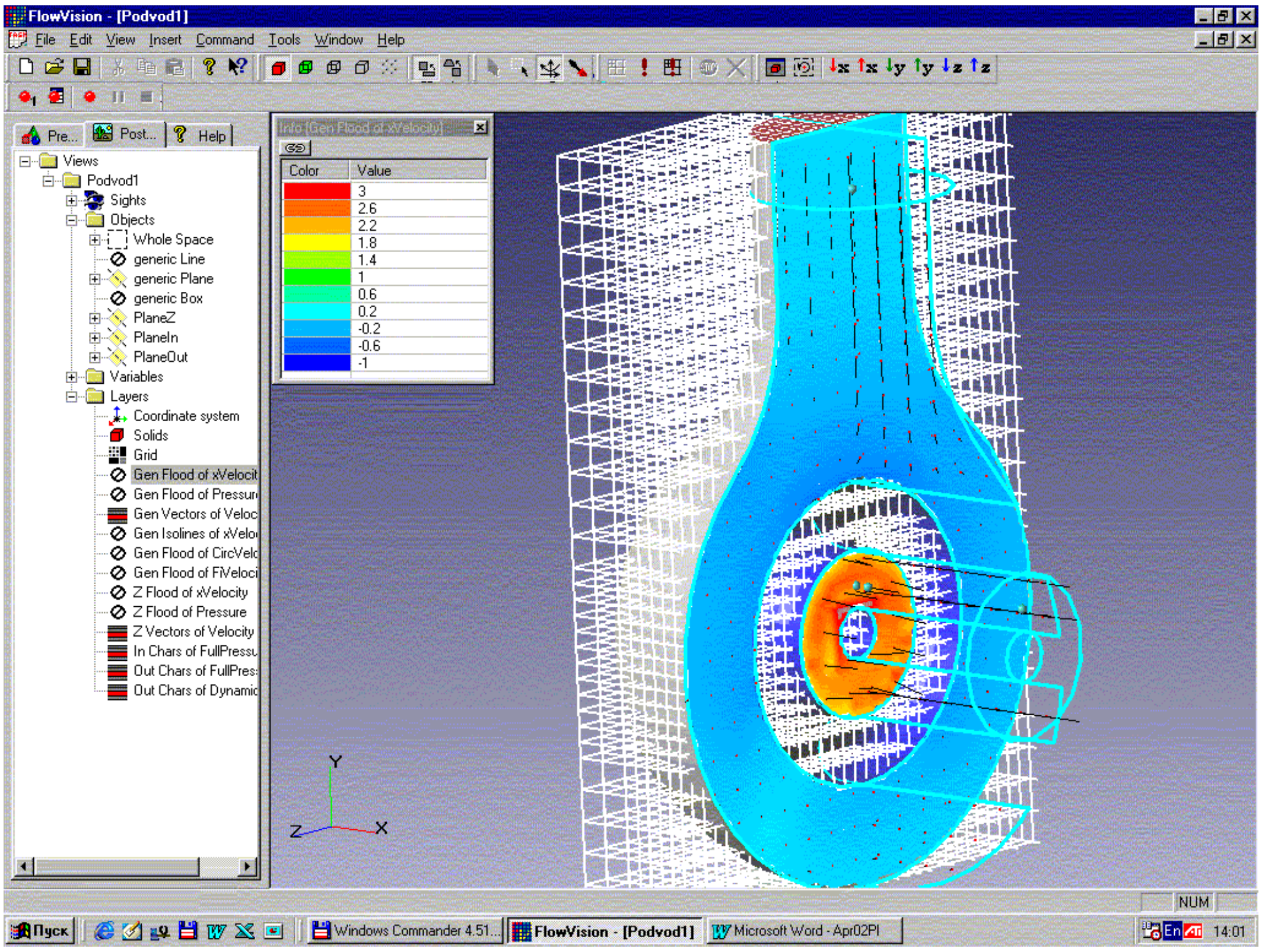

**Figure 1.** Window of the software tool FlowVision. In the window, for a certain problem, the geometrical model, post-processing layers and computational grid are shown.

With this approach, the precision of numerical solution inevitably degrades due to large divergence between grid lines and fluid flow lines. This deficiency is prevented by application of a special skewed scheme developed by authors for approximating of higher-order derivatives (Aksenov et al., 1996).

In this paper, we use only the model of turbulent flow of incompressible fluid that includes time-dependent Reynolds equations projected to each of three Cartesian axes, the continuity equation and the equations of standard $k-\varepsilon$ turbulence model, the only turbulent model implemented in FlowVision nowadays.

We should note, there are a number of other flow models implemented in FlowVision (model of laminar flow, compressible flow, flow with free surface, flow with combustion, etc.) that we don't use in this paper. More detailed information concerning computation of such flows is presented in the user manual and tutorial supplied together with FlowVision.

Discretization of the model flow equations is performed using control volume method. Algorithm of numerical solution of equations has much in common with the known scheme SIMPLE described in the book of Patankar (1980). However, some numerical techniques (special skewed scheme for approximating of derivatives, algorithm for removing of small cells) are used by the developers of FlowVision for the first time.

Details of the model equations and numerical algorithm implemented in FlowVision are described in the papers of Aksenov et al. (1993, 1996).

Solution process starts from a certain initial approximation imposed as source data. Transient solution is obtained without any simplifications. As a result of each global iteration, after elapsing of corresponding time step, new values of velocities and pressure are obtained. Stationary solution, if exists, is reached after completion of a large enough number of iterations which correspond to large enough period of time.

Convenient postprocessor interface permits to see at the screen both flow pattern and its integral parameters (e.g., loss factor between specified cross-sections). The results are updated after each global iteration (as flow pattern changes with elapsing of time). The possibility of saving the intermediate results as text files is available (which can be then processed, e.g., with Microsoft Excel), allowing for monitoring the convergence process, and obtaining the dependence of required parameters on time for transient flows.

## 3. Results

### *3.1. Flow at the initial section of a cylindrical pipe*

At the entrance to the pipe of $D_0$ in diameter, flow velocity is constant through the cross-section. As the flow develops under action of friction forces, velocity distribution approaches the logarithmic law. Experimental data are taken from Barbin et al. (1963), Reynolds number computed by inlet diameter was 388000. Comparison of velocity distributions is performed at several cross-sections (Fig. 2). For this simple flow, computed and experimental velocity distributions coincide well at intermediate cross-sections, and at the last cross-section the results somewhat diverge. Maximum velocity at the channel axis, according to experiment, was $u / U_0 = 1.22$, according to computation – 1.195. More exact results for this flow were obtained using much simpler algorithm (Kochevsky, 2001) that took into account axial symmetry and parabolic character of flow.

In order to check grid independence of solution, the computations were performed for several grids of different thickness. Taking into account good coincidence of results obtained with grids 10 x 10 x 37 and 14 x 14 x 49 (Fig. 2), further thickening of grid for this problem seems to be not necessary. Time for flow computation with the grid 14 x 14 x 49 using PC with processor Pentium III 500 and 64 MB RAM was about half an hour per 50 global iterations.

As an indicator of convergence process, maximum velocity at the last cross-section is convenient to use. Figure 3 presents variation of this velocity with number of global iterations. As it is seen, after 25 iterations variations of this velocity don't exceed 1%. According to recommendations of the user guide, time step corresponding to one global iteration for this and subsequent problems was chosen to be approximately equal to one tenth of time required for a fluid particle to pass through the domain, from the inlet to the outlet.

Loss factor along the length from the medium till the end of the pipe, computed with FlowVision, was 0.292, and according to the formula of Filonenko – Altshul (Idelchik, 1975, Eq. (2.5)) for developed flow in a hydraulically smooth pipe it should equal to 0.238. The correspondence of results is acceptable, taking into account that in our case the flow is not yet developed, and restructuring of flow implies additional losses.

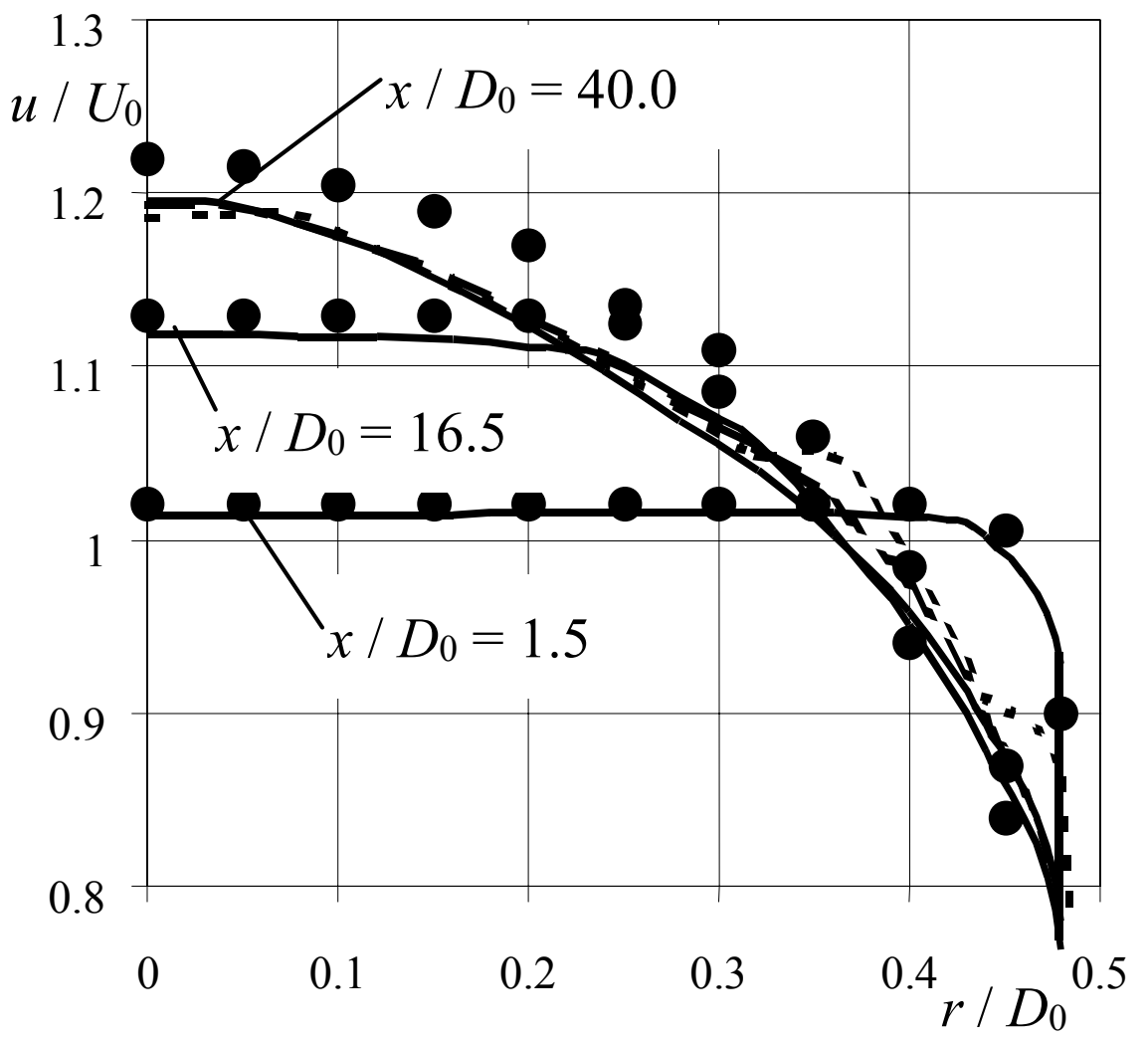

**Figure 2.** Velocity distributions of turbulent flow at the initial part of a cylindrical pipe: solid line – computations with the grid 14 x 14 x 49, large-dotted line – with the grid 10 x 10 x 37, small-dotted line – 6 x 6 x 25

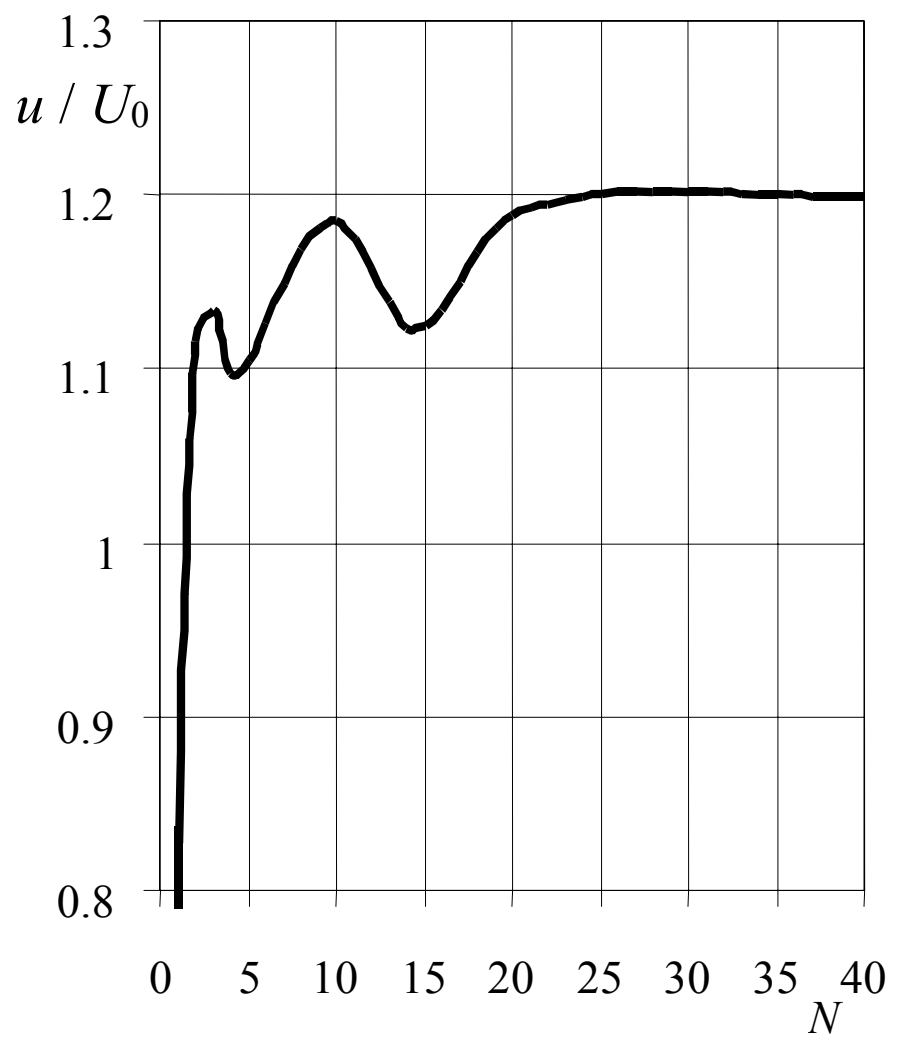

**Figure 3.** Dependence of velocity at the channel axis of the last cross-section on the number of global iterations, computations with the grid 14 x 14 x 49

### *3.2. Flow in diffusers of large internal angles*

An important advantage of the software tool FlowVision is possibility to compute successfully flows with symmetrical computational domain and symmetrical boundary conditions, where the flow is not symmetrical and, moreover, transient.

So, in a conical diffuser with large internal angle the flow separates and attaches to part of the wall in random way. The position of separation zone varies randomly along circumference of diffuser (Idelchik, 1975). This known physical phenomenon is reproduced by computation (Fig. 4). Figure 5 presents corresponding loss factors obtained by computation in FlowVision and taken using interpolation from that reference book.

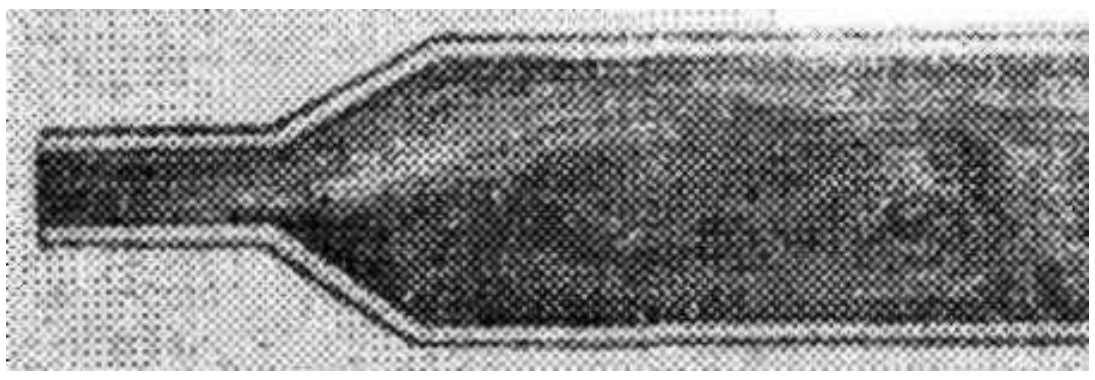

a) experiment, Idelchik (1975)

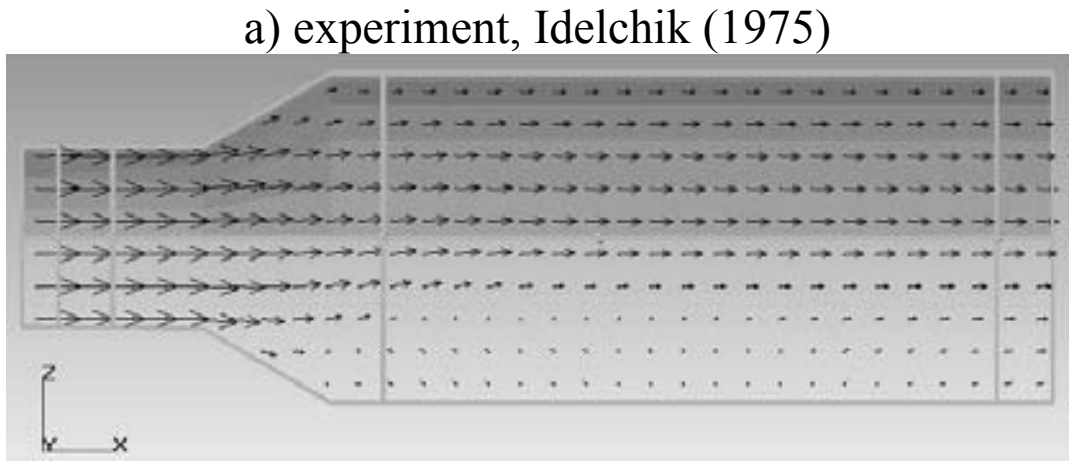

b) computation

**Figure 4.** Instant flow pattern in a conical diffuser with area ratio 3.3 and internal angle 60°

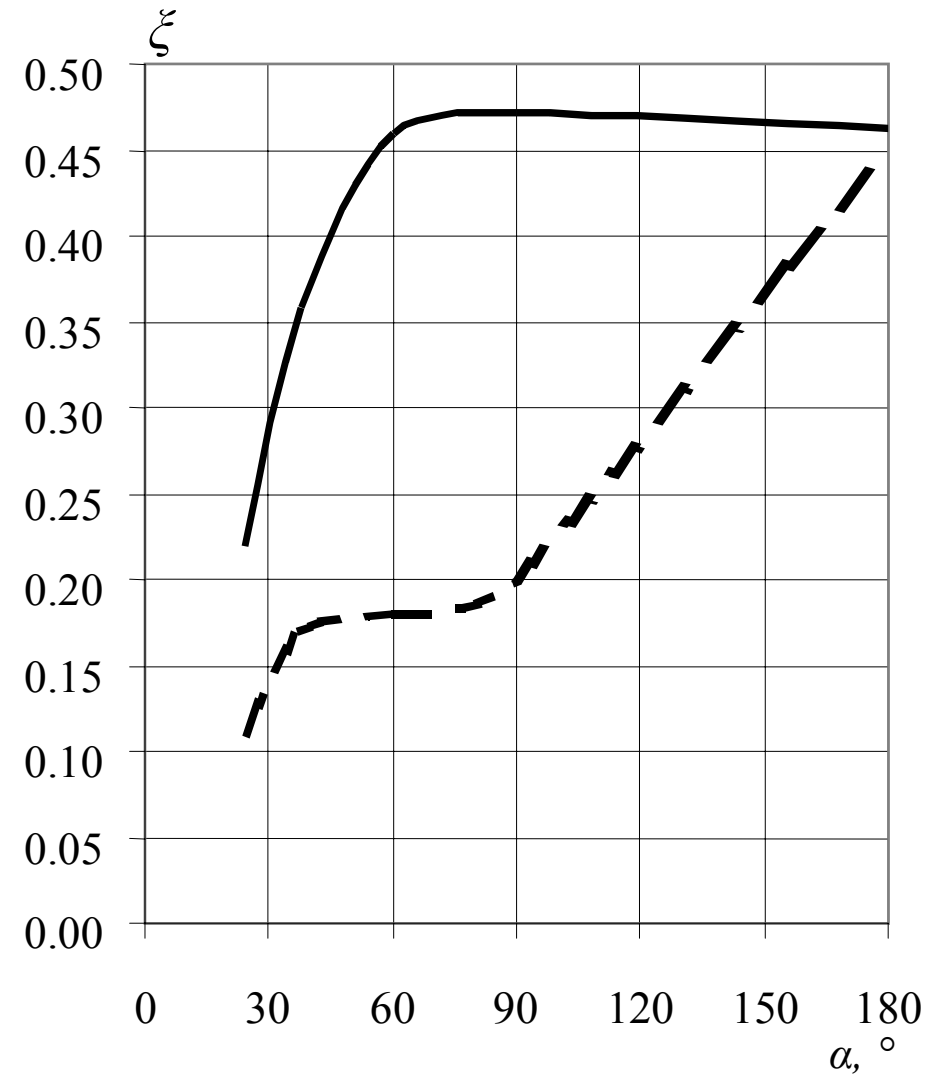

**Figure 5.** Dependence of loss factor on internal angle of a conical diffuser with area ratio 3.3: solid line – experiment, Idelchik (1975), dotted line – computation

Note that for this flow the solution process does not converge to the stationary solution, and the computed loss factors presented at Fig. 5 are approximate time-averaged values. Considering that loss factor in a diffuser can substantially depend on a number of factors that are difficult to take into account, the correspondence of results can be regarded as satisfactory.

***3.3. Solid body swirled flow in a conical diffuser***

Such swirl can be obtained if the flow enters the diffuser passing through a section filled with straws or honeycomb, and this section rotates relative to the diffuser axis using an external drive. For performing the comparison, experimental results of Clausen et al. (1993) are taken. Area ratio of the diffuser (ratio of areas of outlet and inlet cross-sections) was 2.84, internal angle – 20°, Reynolds number computed by inlet diameter – 212000. As the source data, distributions of axial and circumferential velocity at the inlet cross-section are specified.

Here, the following physical effect occurs: as swirl weakens downstream (due to expansion of flow and action of friction forces), the flow is pressed to the periphery. In the experiment, the swirl was so strong that axial velocity at the diffuser axis dropped almost till zero (when the inlet swirl is still stronger, reverse flow along the diffuser axis is observed). It makes impossible to use parabolic method of computation, like that in the paper of Kochevsky (2001).

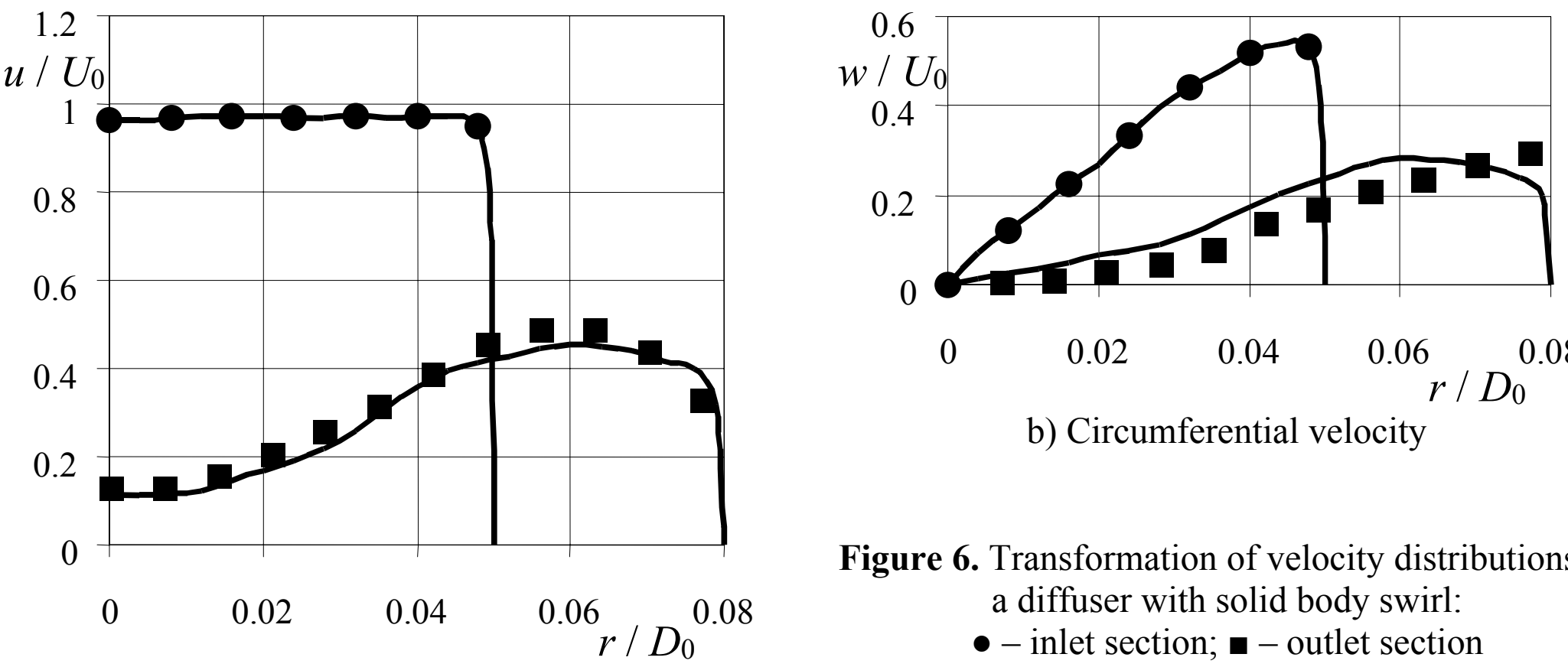

a) Axial velocity

b) Circumferential velocity

**Figure 6.** Transformation of velocity distributions in a diffuser with solid body swirl: ● – inlet section; ■ – outlet section

Figure 6 presents distributions of axial and circumferential velocity at the inlet and outlet cross-sections of the diffuser. The correspondence of results computed in FlowVision with experimental results is good for this flow. Similar flow was investigated in Armfield et al. (1990) with the same turbulence model, where obtained correspondence of computed and experimental results was as good as here.

***3.4. Free vortex swirled flow in a conical diffuser***

Such swirl is added to flow, for example, when flow passes between guide vanes. Near channel axis, the circumferential velocity decreases linearly till zero. For performing the comparison, experimental results of Senoo et al. (1978) are taken. Area ratio of the diffuser was 4.0, internal angle – 12°, Reynolds number computed by inlet diameter – about 200000.

In this experiment, the swirl intensity also was so large that the axial velocity almost reached zero at the diffuser axis. Figure 7 presents distributions of axial and circumferential velocity at the inlet and outlet cross-sections. Again, quite good correspondence of computed and experimental results is observed.

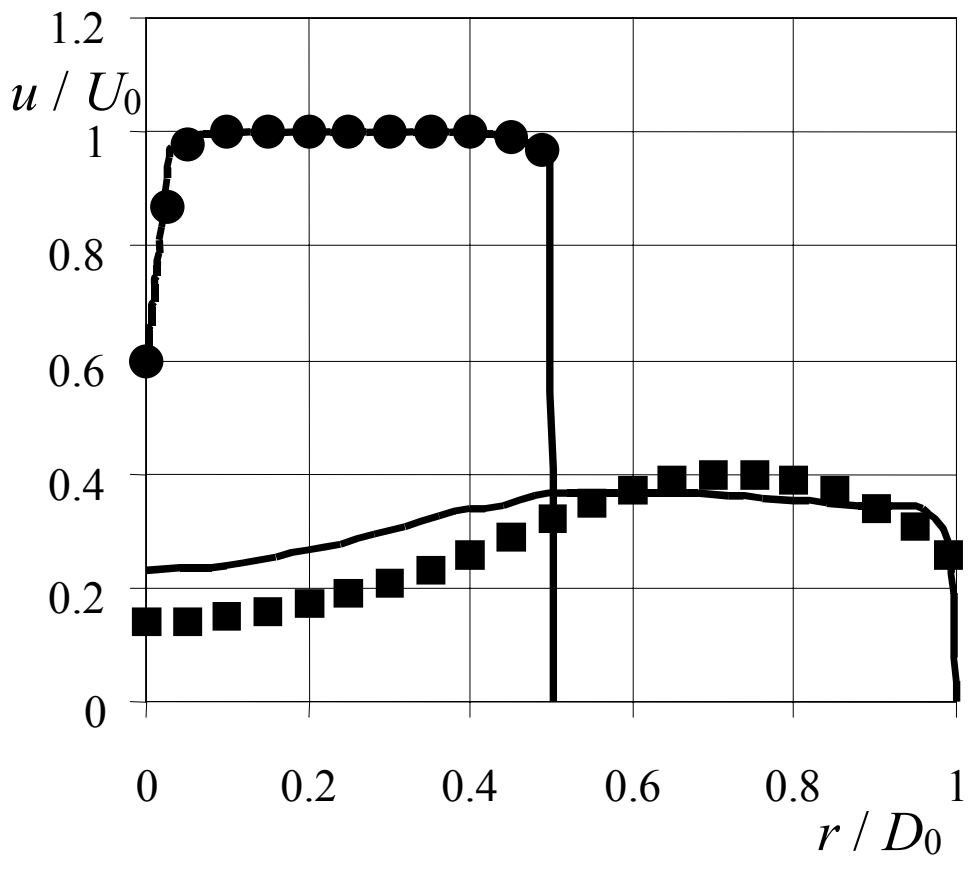

a) Axial velocity

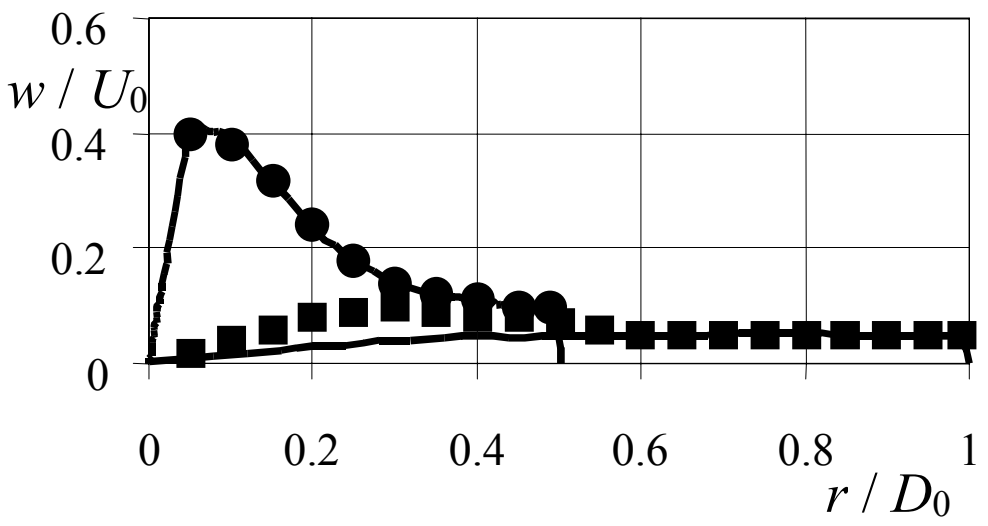

b) Circumferential velocity

**Figure 7.** Transformation of velocity distributions in a diffuser with free vortex swirl:
● – inlet section; ■ – outlet section

### *3.5. Flow in a bend of square section with 180-degree flow turn*

For comparison, experimental results of Chang et al. (1983) are used. The considered channel is presented at Fig. 8. The flow is symmetrical relative to medium plane of the channel, thus, in order to spare computational resources, only half of the channel was used as the computational domain. Reynolds number computed by hydraulic diameter was 58000.

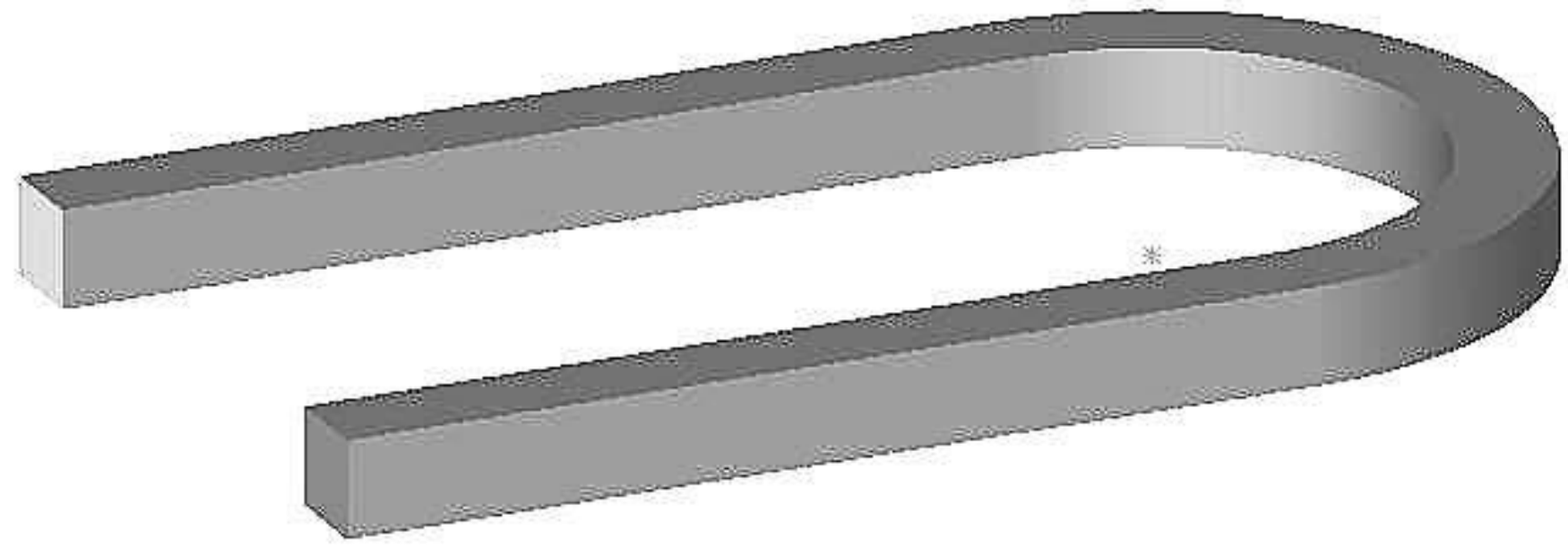

**Figure 8.** Geometrical configuration of the channel

A physical effect peculiar to this flow is the twin vortex occurring as the flow turns (Fig. 9). Fluid moves to the external radius of bend along the symmetry plane of the channel and returns back along the side walls. Figure 10 presents distributions of axial velocity at several intermediate cross-sections of bend ($H$ – side of a cross-section of the channel). The distributions are taken at the plane equidistant from the symmetry plane and lower wall of the channel.

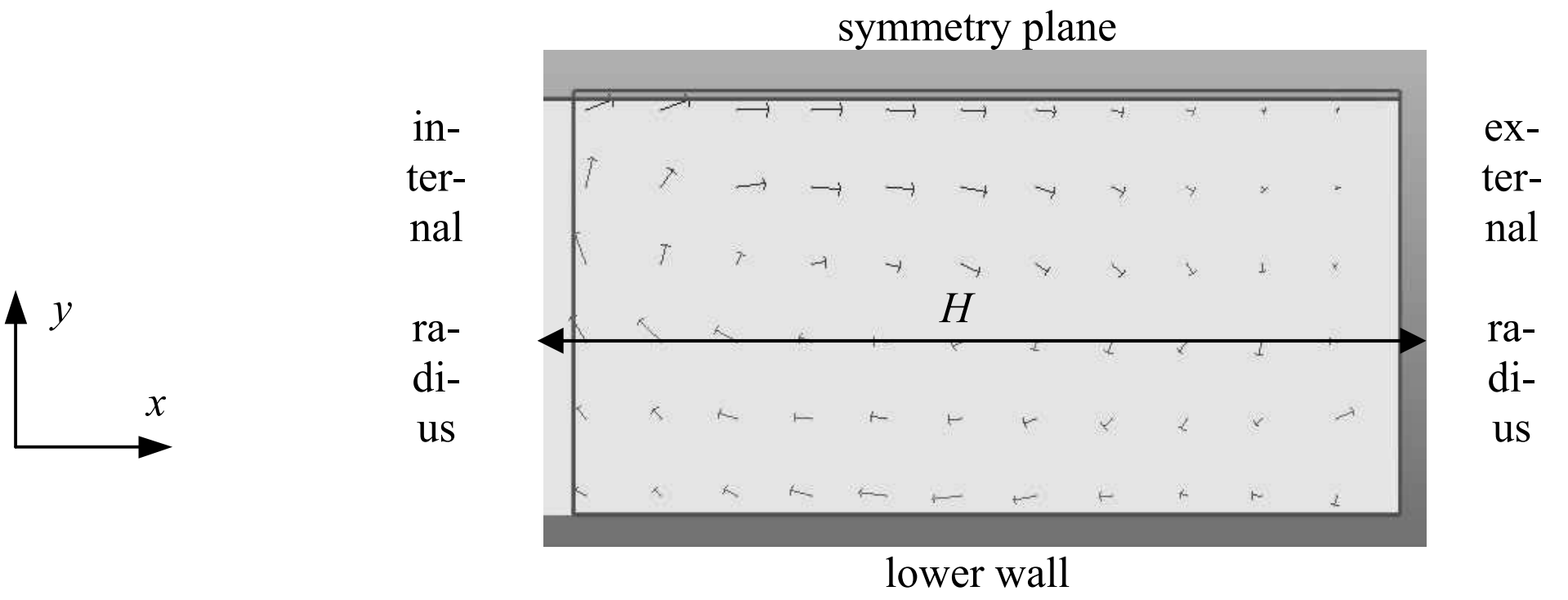

**Figure 9.** Velocity vectors at the cross-section corresponding to 90-degree flow turn

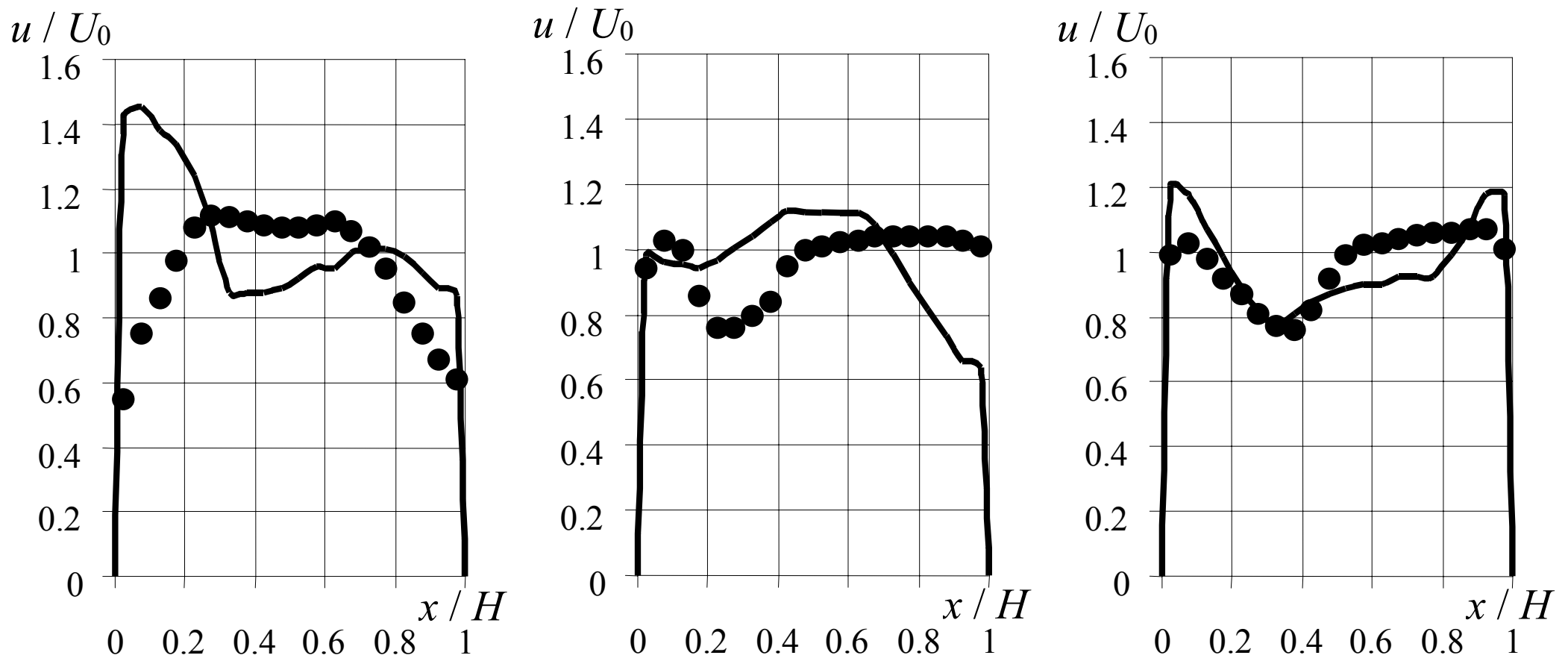

a) Angle of turn 45° b) Angle of turn 90° c) Angle of turn 130°

**Figure 10.** Axial velocity distributions at the channel turn ($y$ / $H$ = 0.25)

As one can see, the computed and experimental results look similar, however, significant quantitative divergence is observed. Both height of peaks near channel walls and position and depth of pit at the middle obtained by computation don't coincide with those observed experimentally.

This flow was investigated numerically also in the paper of Choi et al. (1989) using a similar computational algorithm but more complex turbulence models. The conclusion was made that for this flow, the standard $k - \varepsilon$ turbulence model gives somewhat inaccurate solution, more precise results were obtained, in particular, using the Reynolds algebraic stress model. Another reason for difference of results at Fig. 10 is seemingly application of Cartesian computational grid that degrades precision of discretization of derivatives.

### *3.6. Flow in a rotating channel*

For comparison, experimental results of Moore (1973) are used. Geometrical configuration of the channel is presented at Fig. 11. As one can see, it resembles in many aspects the geometrical configuration of a blade-to-blade channel of a radial-flow impeller with radial blades. The channel rotates around the axis of inlet cylindrical section. The flow enters the channel along its rotation axis, then turns in radial direction and expands. Outlet section of the channel is open to atmosphere.

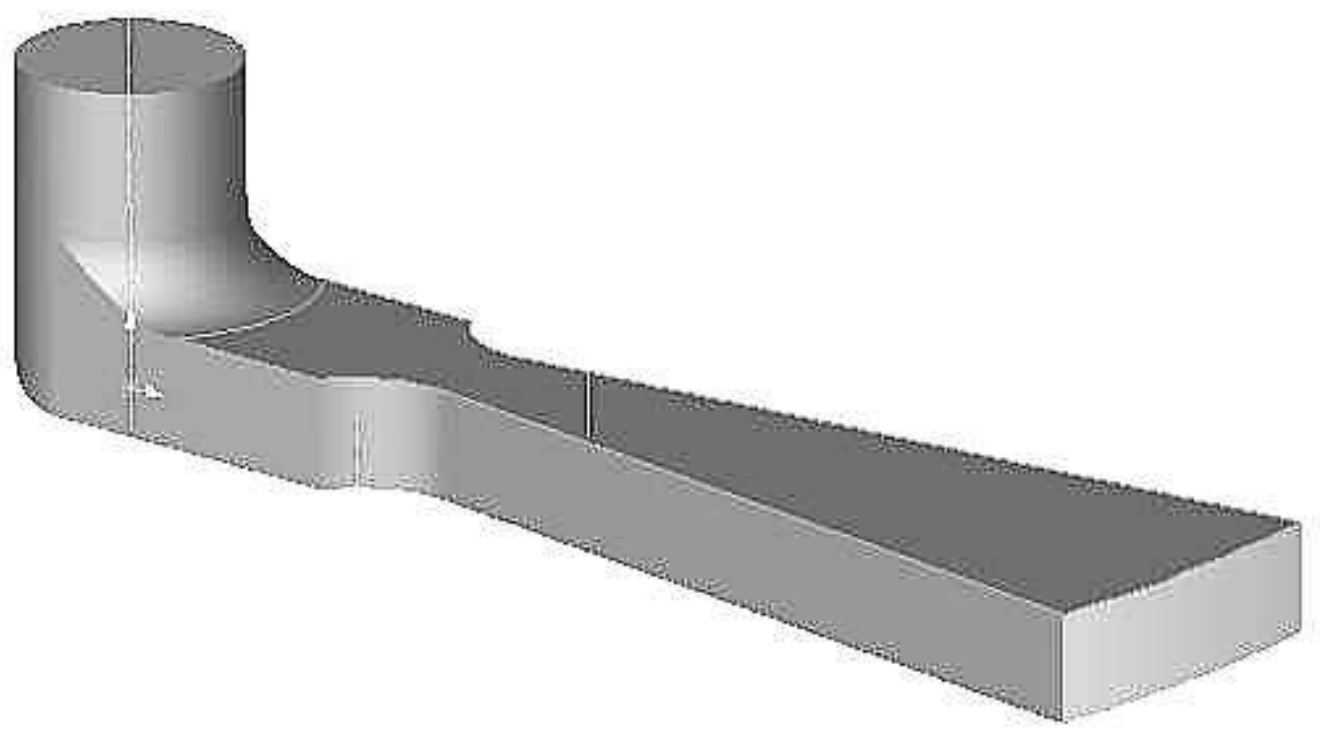

**Figure 11.** Geometrical configuration of the channel

The flow enters the channel through a stabilizing device ensuring the through-flow velocity is constant through the inlet cross-section. Then channel walls gradually involve the flow in rotation. Rotational speed of the channel was 206 rpm. Upstream of stabilizing device, a blower was provided, allowing for supply of air to the channel with different flow rate.

This flow is featured with the following physical effects.

- As distance from the rotation axis increases, static pressure grows rapidly due to action of centrifugal force. Additionally, it grows due to expansion of the channel.
- The pressure $p$ is larger at the pressure side than at the suction side. As it follows from the analysis of potential flow (not regarding viscous effects), the pressure decreases from the pressure side to the suction side, according to the linear law $\partial p / \rho \, \partial y = -2 \, \omega \, u$, where $\omega$ is angular rotational speed of the channel.
- As it follows from the analysis of potential flow, through-flow velocity $u$ increases in every cross-section from the pressure side to the suction side, according to the law $\partial u / \partial y = 2 \, \omega$ (Fig. 12).
- At low flow rate, the flow in the channel is pressed to the suction side along the whole length of the channel. Thus, at the pressure side, the reverse flow is observed when the distance from the rotation axis is large enough (Fig. 12).

All these effects are well reproduced by computation.

- According to the experiment, when the flow rate is large enough, the flow at some distance from the rotation axis separates from the suction side and is pressed to the pressure side. The reason for this is seemingly in rapid growth of pressure along the suction side as the channel expands (Fig. 13). Just near this wall the flow is inclined to separation. At low flow rate this separation does not occur, because earlier the flow separates from the pressure side.

This separation occurs also in computation (Fig. 14), though the computation predicts separation also from the pressure side that was not observed experimentally.

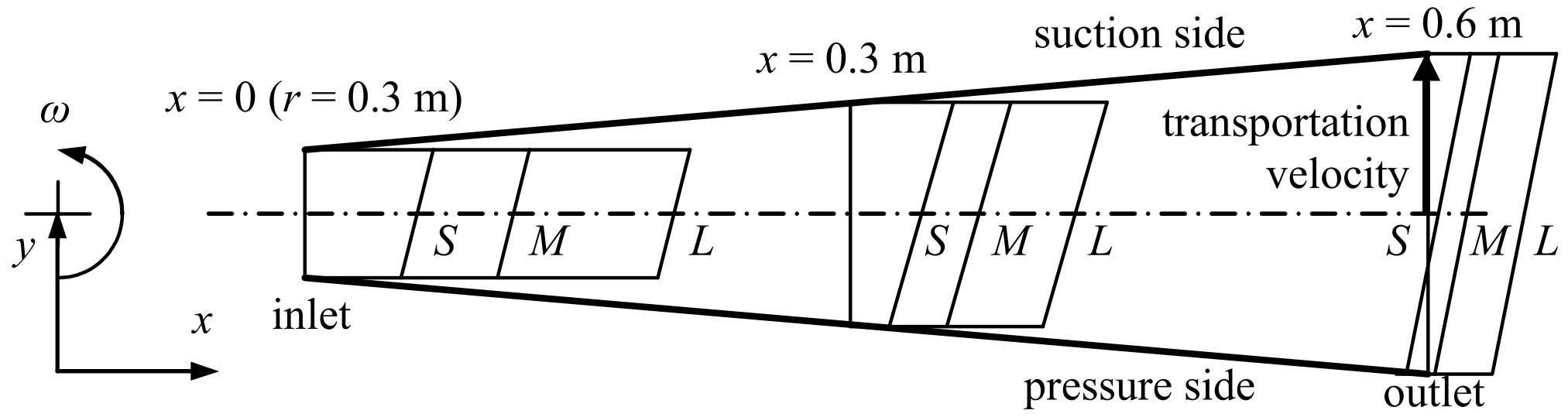

**Figure 12.** Velocity distribution in the channel according to the potential flow analysis: S – low flow rate (average velocity at the outlet cross-section of the channel is 2.8 m/s), M – medium flow rate (5.3 m/s), L – high flow rate (11.0 m/s)

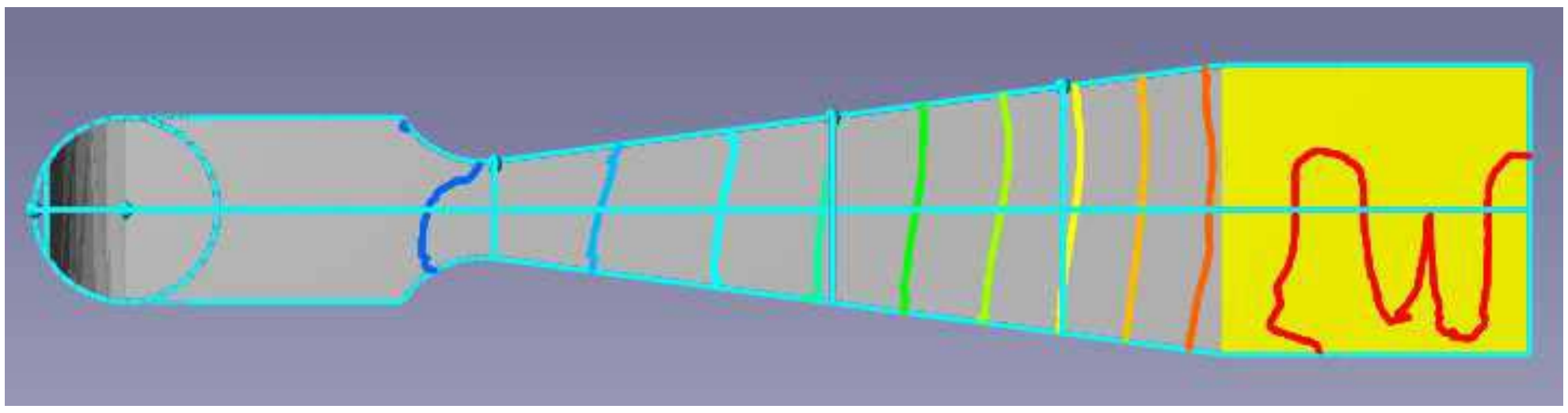

**Figure 13.** Isolines of static pressure at the middle-height cross-section of the channel at medium flow rate, computation

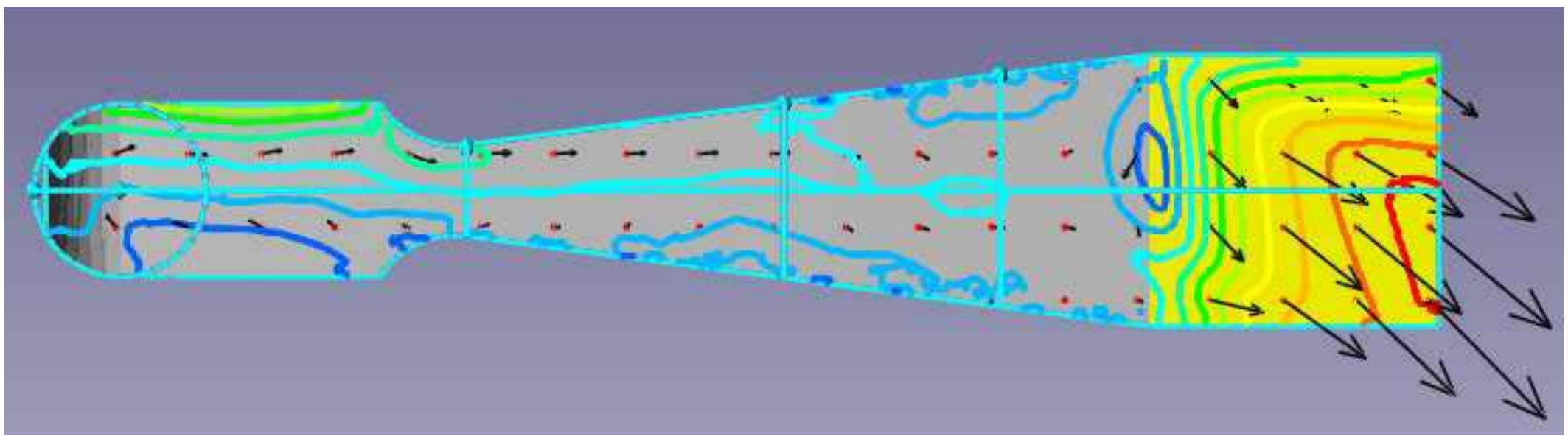

**Figure 14.** Isolines and velocity vectors in relative motion at medium flow rate, computation.
At the middle part of the channel, isolines restrict zones of reverse flow near pressure side and suction side

This rather simple experiment is a good illustration explaining why location of separation zones in blade-to-blade channels of a radial-flow impeller depends significantly on the flow rate.

For performing computation in FlowVision, we have used the computational grid containing 46 x 12 x 15 nodes, with one additional level of splitting for cells crossing solid surfaces. The computation was performed in relative motion. Here, as the inlet boundary condition, in addition to through-flow velocity, solid body swirl was imposed. The rotational speed of the swirl was – 206 rpm (opposite to the direction of channel rotation). In order to simulate properly the flow at the outlet region, the computational domain included also the space downstream of the channel, where the flow went to the atmosphere (Fig. 13 – 15). Figure 15 shows vectors of flow velocity: at the inlet region – in relative motion, at the outlet region – in absolute motion.

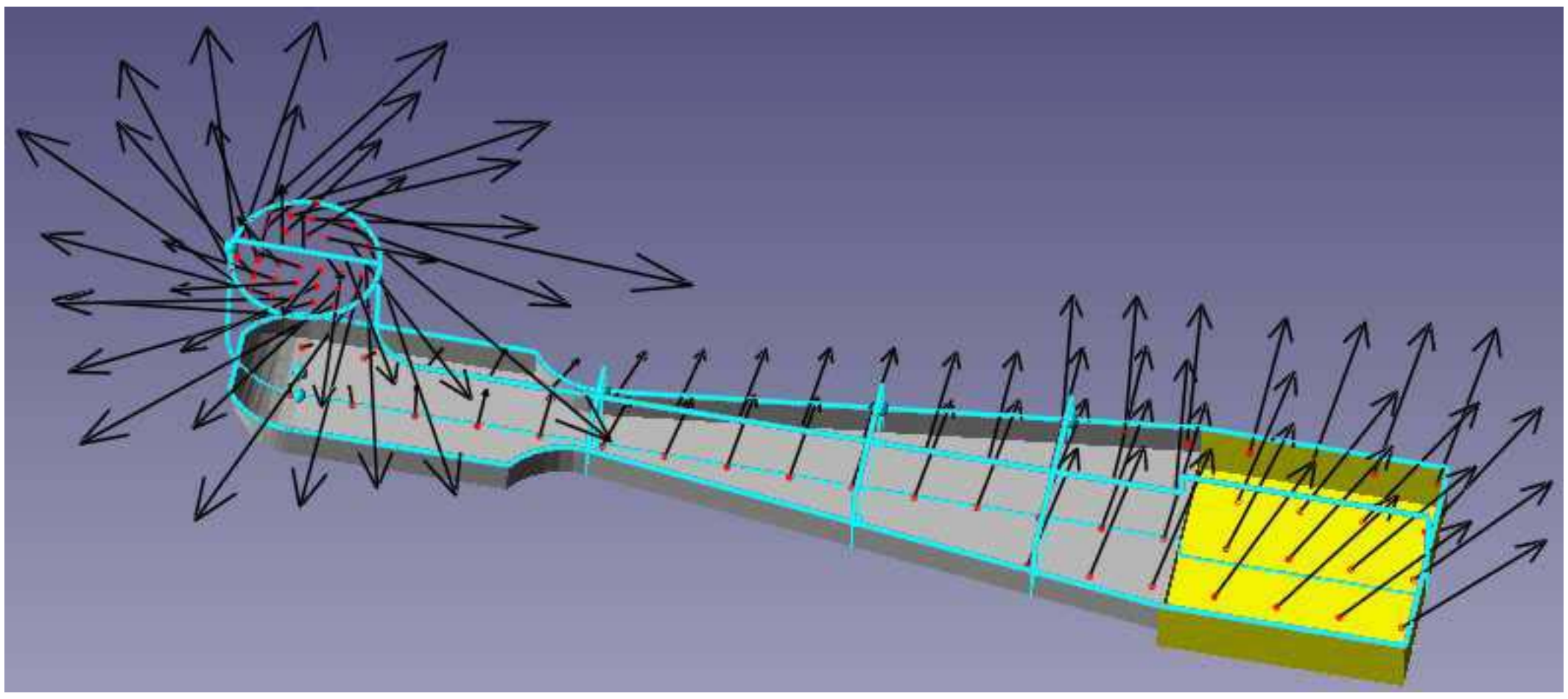

**Figure 15.** Velocity vectors of flow in relative motion (at the channel inlet)
and in absolute motion (inside the channel and at the channel outlet) at medium flow rate

Note that in the paper of Majumdar et al. (1977) this flow was computed using a sweep scheme, with the same $k-\varepsilon$ turbulence model. The separation of flow from the suction side was not observed in that paper, thus demonstrating that Reynolds equations should not be simplified by order-of-magnitude analysis for simulation of this flow.

Figure 16 presents computed in FlowVision and experimental velocity distributions at the cross-section before channel outlet, at different flow rates. As can be seen, the listed experimental physical effects are reflected by computation, however, significant discrepancy of computational and experimental results is available. The reasons for this discrepancy are seemingly the same as for the previous flow.

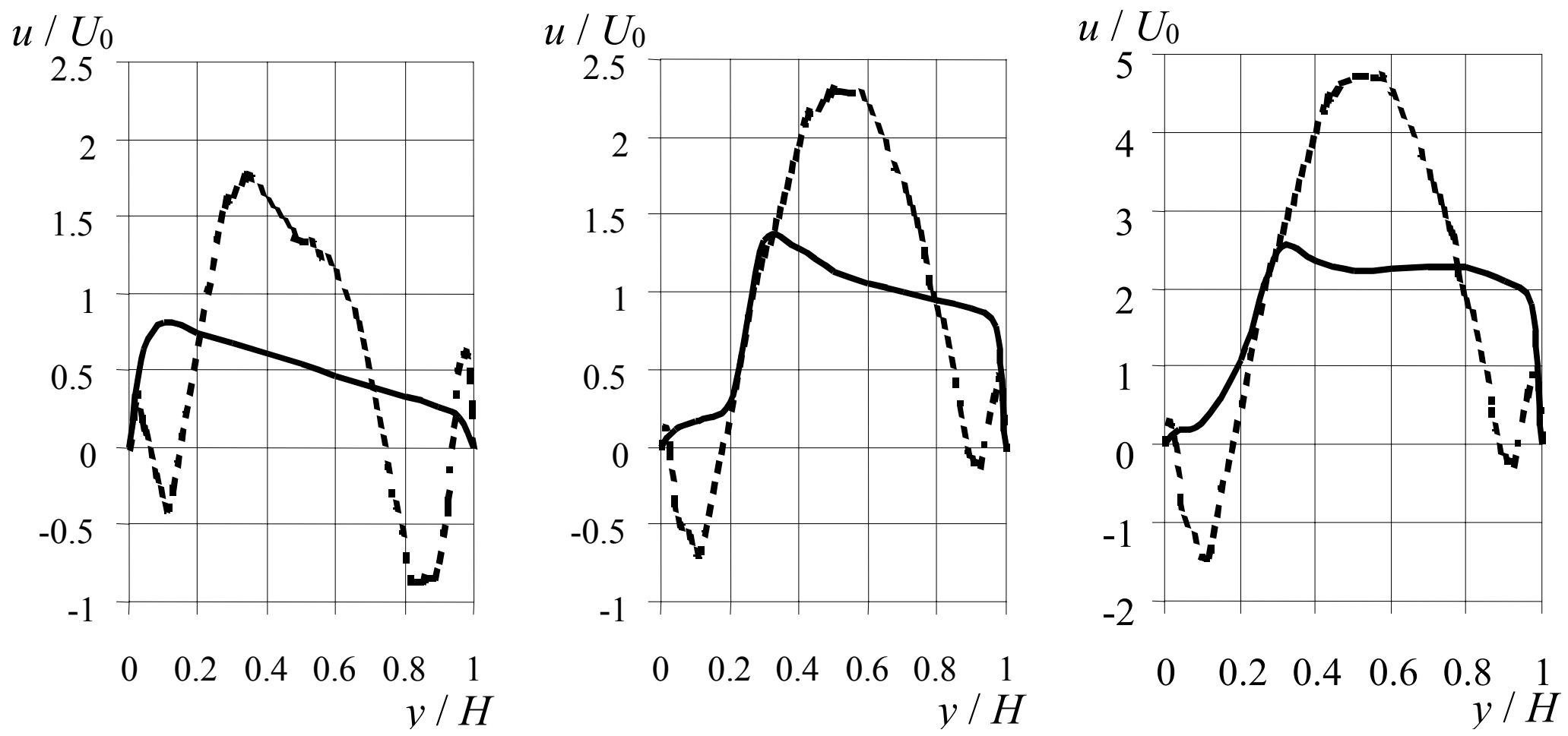

a) Low flow rate b) Medium flow rate c) High flow rate

**Figure 16.** Distributions of through-flow velocity at the cross-section $x$ = 0.46 m: solid line – experiment, Moore (1973), dotted line – computation. Left-hand side is the suction side. $U_0$ – average through-flow velocity at the channel inlet, at the medium flow rate, $H$ – height of the channel cross-section

## 4. Conclusion

For the flows presented here, we have obtained good qualitative and at least satisfactory quantitative correspondence of results obtained using the CFD software tool FlowVision with experimental results. For types of flows presented here, we should recognize the approach implemented in FlowVision to be rather good for numerical simulations.

### *Acknowledgements*

The present research was conducted under leadership of Asst. Prof. A.A. Yevtushenko and under support of the collective of the department of fluid mechanics.